\documentclass[12pt]{article}
\title{Faster Than Light?}
\author{Robert Geroch\thanks{E-mail:  geroch@uchicago.edu}
\\Enrico Fermi Institute\\5640 Ellis Ave,
Chicago, IL, 60637, USA}

\begin{document}
\maketitle

\noindent {\bf \large Abstract.} It is argued that special
relativity remains a viable physical theory even when there
is permitted signals traveling faster than light. 
\\

\section{Introduction}

The view is widely held that the speed of light, $c$, plays a fundamental
role in physics:  It is the universal 
upper limit to all signal speeds that can be achieved, by any process 
whatever, in the physical world.   This view comes to us, of course, 
from special relativity.  Indeed, it is normally taken as a fundamental
tenant of that theory --- be it as an ``axiom", on which the theory is 
based; or as a consequence of
other ingredients of the theory.  But our confidence in this special
role for the speed $c$ is based
on more than merely its status within special relativity.
There are also solid physical arguments to support this position.

One such argument is the following.  Try to concoct a 
physical mechanism that could be used to generate a superluminal signal.  
For example, one might, by applying various forces to a particle,
attempt to accelerate it up to a speed exceeding $c$.  Such
a particle, once so accelerated,  could then be used to transmit 
signals.  But such mechanisms seem invariably to
fail.  In this example of a particle, for instance, we must contend
with the formula, $m = m_o/\sqrt{1-v^2/c^2}$, for the
mass-increase of a particle with its speed $v$.   This formula 
guarantees that no application of any finite force for any
finite amount of time will ever achieve $v > c$.  It appears, then,
that Nature conspires to prevent the manufacture of
superluminal signals.

A very different kind of argument is that of the so-called grandfather
paradox.  Suppose that one had produced a mechanism for sending a signal
faster than light.   For example, suppose that one had constructed
a long pipe containing a fluid whose sound-speed 
exceeds $c$.  This means that the world-line of a sound pulse
sent down this pipe would be spacelike.  Assuming that
Lorentz invariance holds for such fluid-filled pipes,
it follows:  For a pipe moving, 
along its length, with a speed comparable with (but
less than) $c$, the world-line of a sound pulse going
down that pipe in the direction of its motion would actually
go backward in time (as measured by the observer
with respect to which the pipe is moving).  Now let our
observer arrange two
such pipes, parallel to each other and close together; with
one at rest and the other moving by at high speed along its length.  
Send a sound pulse down the rest pipe, and, after that signal
has traveled for a while, have it initiate a second sound pulse 
sent back to the observer along the moving
pipe.  The result of this arrangement would be to send 
a signal into this observer's
own past.  By using a sequence of such arrangements, our observer
could send a signal very far into his own past.  Finally,  
this observer
could use such a signal to change something in the past --- e.g., to
have his own grandfather killed while still a child.  But now we
have a paradox:  If the observer's grandfather was killed as
a child, then the observer would never have come into being,
and so would never have been able to construct this machine for the murder
of his grandfather!  It appears to be difficult to resurrect
``physics" (at least in any form that we are familiar with) 
in the presence of superluminal signals.

We shall argue here that, all this evidence notwithstanding,
special relativity need {\em not} be construed as prohibiting
superluminal signals.  Relativity theory with such signals
permitted, we shall argue, is as viable and physically acceptable 
as relativity theory without.  
We suggest that a universal limitation on signal speeds
need not be taken as any fundamental principle of physics. 
Rather, the whole idea of such a limit has more to do with 
history and with the types of interactions to which we are commonly exposed.
We emphasize that we are {\em not} suggesting here that some
new theory be introduced to replace special relativity; nor,
indeed, that any of the basic structural components of the
theory of relativity be changed.  What is to be changed is merely 
our perspective on relativity theory.

This claim is based on some features of a class of partial
differential equations called symmetric hyperbolic systems.
This subject is reviewed briefly in Sect. 2.  We return to 
special relativity in Sect. 3.

\section{Symmetric Hyperbolic Systems}

In this section, we review the properties of a certain class 
of systems of partial differential equations \cite{AR}, \cite{KOF},
\cite{FJ}, \cite{PDL}.  Our treatment follows \cite{PDE}.

Denote by $M$ the 4-dimensional manifold of space-time events.
Consider a first-order, quasilinear system of partial differential
equations on this manifold.  That is, consider a system of equations
of the form 
\begin{equation}
k^{Aa}{}_{\alpha}\ (\nabla_a \phi^{\alpha}) = j^A.
\end{equation}
Here, $\phi^{\alpha}$ represents the fields of the system, where
the index ``$\alpha$" runs through field-space\footnote{A more elegant
formulation of this scheme is the following. 
Introduce, at each point of $M$, the manifold of ``all possible field 
values" at that point.  Then the set of all such field-values at all 
points of $M$ becomes a fibre bundle over $M$.  The field-configuration
of the system, $\phi$, is then represented as
a smooth cross-section of this bundle.  This formulation, among other
things, makes it clear
that how the fields are represented --- e.g., as densities, or 
contravariant tensors,
or whatever --- plays no role in what follows.}.  Any algebraic
conditions on the fields --- be they equalities or inequalities ---
are to be reflected in the construction of the $\phi^{\alpha}$:  
There are no ``algebraic constraints" on the fields in this formulation. 
The expression ``$(\nabla_a \phi^{\alpha})$" in (1) is the system of
first derivatives of these fields\footnote{In the more elegant
formulation, we fix any smooth connection on the field-bundle, and then
use this connection to take the derivative, in (1), of the cross-section.
Which particular connection is chosen
plays no essential role in what follows.}, where the 
index ``$a$" labels tensors in $M$.   Eqn. (1), then, requires that 
certain expressions linear in these derivatives vanish.   
The coefficients in these linear
combinations, $k^{Aa}{}_{\alpha}$ and $j^A$, are any
fixed smooth functions of the fields $\phi^{\alpha}$ (but not of their
derivatives) and the point of the
underlying manifold $M$.  Eqn. (1) is to be imposed at each point
of $M$.  The free index, ``$A$", in (1) lives in the vector space of
equations, i.e., there is one equation for each ``value" of $A$.
This system is called first-order, since
it involves at most first derivatives of the fields $\phi^{\alpha}$; and 
quasilinear, since the equations themselves are linear in those first
derivatives.  We demand at the outset that {\em all} first-order equations
on the fields --- even
any that might be derived from the other equations by taking derivatives ---
have been included in (1).

The vast majority of systems of equations that describe physical
systems can be cast into this form. For electromagnetism,
for example, the field $\phi$ is a skew, second-rank tensor field
$F_{ab}$.   And Eqn. (1), of course, is Maxwell's equations: 
\begin{equation}
\nabla^b F_{ab} = 0,
\end{equation}\begin{equation}
\nabla_{[a}F_{bc]} = 0.
\end{equation}
Here, the ``field index", $\alpha$, runs over the six dimensions
of field-space, while the ``equation index", $A$, runs over 
the 8-dimensional vector space of equations (2)-(3).  A second
example is that of a simple perfect fluid.  The fields in this
case consist of a unit timelike 
vector $u^a$ (the fluid 4-velocity) and a positive scalar $\rho$
(the mass density); and Eqn. (1) is the system of fluid equations:
\begin{equation}
(\rho + p)u^m\nabla_m u^a + (g^{ab} + u^au^b)\nabla_b p = 0,
\end{equation}\begin{equation}
u^a\nabla_a \rho + (\rho +  p)\nabla_au^a = 0.
\end{equation}
In these equations, $p$ is some fixed nonnegative function of $\rho$, the function 
of state, which characterizes the fluid under consideration.  
Here, index ``$\alpha$" runs over
the 4-dimensional space of pairs $(u^a, \rho)$ with $u^a$ unit
and $\rho > 0$; while ``$A$" runs over the 4-dimensional vector
space of pairs $(w^a, w)$, where $w^a$ is orthogonal to $u^a$,
and $w$ is a number.  The coefficients $k^{Aa}{}_{\alpha}$ and $j^A$ in these
examples are read off immediately from the equations.
Similar remarks apply to other systems, for example, those of 
neutrino or Dirac fields, of spin-$s$ fields, of stressed
solids, of more complex
fluids (e.g., those with a higher-dimensional manifold of local 
fluid states), of fluids manifesting dissipative effects 
(viscosity, thermal conductivity), etc.  
Second- and higher-order equations are reduced to first-order form
by introducing new fields to represent the lower derivatives of the
original fields.  Thus, for example, the wave equation
is represented by a pair of fields, $(\psi, v_a)$, with equations
$\nabla_a\psi = v_a, \nabla_{[a}v_{b]} = 0$, and $\nabla^a v_a = 0$.
[Note that we include in our system the second of these equations, 
even though it follows from the first by taking the curl.]
And, finally, Einstein's equation can be cast into the form of
Eqn. (1).  The fields in this case are the space-time metric $g_{ab}$ 
and the derivative operator $\nabla_a$; and the equations are
$\nabla_a g_{bc} = 0$ and $G_{ab} = 0$, where $G_{ab}$ is the
Einstein tensor.  Note that these equations are indeed first-order
and quasilinear. 

Consider two systems --- say those of the electromagnetic field and
of a simple fluid.  We can imagine a world in which these two systems coexist
independently of each other.  This situation is described by
fields that merely combine the fields of the individual systems; and  
equations that combine the equations of the individual systems.
Thus, in our example, $\phi^{\alpha}$ for the electromagnetism-fluid system
would consists of a skew tensor $F_{ab}$, a unit timelike $u^a$, and
a positive scalar $\rho$ (10 dimensions of fields);  and the system of
equations would be the total system (2)-(5) (12 dimensions of equations). 
Next, let there be turned on an interaction between these two systems.
Nature, apparently, always ``turns on interactions" in a very
special way:  It is only the $j^A$ of the combined system that is modified
to reflect the interaction, while the $k^{A a}{}_{\alpha}$
remains the same.  Consider, in the example above, the
interaction that results from allowing the fluid to
carry charge.  First introduce the charge-current vector,
$J^a = \mu u^a$ of the fluid, where $\mu$ is a certain function
of $\rho$ so designed that $\nabla_a J^a = 0$ follows already from
Eqn. (5).  Then modify the system (2)-(5) by adding the term
$J_a$ on the right in (2) (The charge-current creates electromagnetic
field.), and adding the term $F^{ab}J_b$ on the right in (4) (The fluid is
subject to the Lorentz force.)  But note that the terms that have been added
to our equations are algebraic in the fields of the combined system,
i.e., that they involve no derivatives of those fields.  In other words,
we have merely modified the $j^A$ of that system.  To take
another example, let us turn on the interaction between a mass-$m$ 
Klein-Gordon
field and gravity.  This is done by inserting the field stress-energy,
$(v^av^b - (1/2)g^{ab}v^sv_s - m^2 g^{ab}\psi^2/2)$, on the right side of 
Einstein's equation.
Again, what is changed is only the $j^A$ of the combined system,
and not $k^{Aa}{}_{\alpha}$.  [Note that we are here making use of
the fact that $v_a = \nabla_a\psi$ has been included among the ``fields" 
for the Klein-Gordon system.]

The view here is that the fields $\phi$, and their equation (1), is 
all there is.  In particular, there is no need for a separate
chapter explaining how each field is to be interpreted.  We ``interpret"
a field exclusively through the physical effects that it produces, 
i.e., by making observations on it.  But an observation on a field, 
in turn, consists of nothing more than the result of turning on an 
interaction, in Eqn. (1), between 
that field and various observer-fields.  Thus, we think of Eqn. (1) 
as a kind of ``theory of everything" (at least, everything
non-quantum-mechanical).  

We now turn to the initial-value formulation a for first-order,
quasilinear, system of partial differential equations.  For
purposes of the present treatment,
we shall ignore two features of these systems --- the possible 
presence of diffeomorphism freedom and of constraints.  These two
features certainly play a significant role in the mathematics
of the general first-order, quasilinear system.   But by ignoring
these two features we shall greatly simplify the present discussion,
while losing nothing of significance.  That is, we are ignoring 
these features merely for convenience. 

A {\em hyperbolization} of the system (1) is a tensor $h_{A\beta}$
(depending, in general, on field-values $\phi^{\alpha}$ and point
of $M$) such that i) the combination $h_{A\beta}k^{Aa}{}_{\alpha}$
is symmetric in the indices $\alpha$ and $\beta$; and ii) for
some covector $n_a$ in $M$, the symmetric tensor 
$n_ah_{A\beta}k^{Aa}{}_{\alpha}$ is positive-definite.  The
equations for the standard systems of physics -- electromagnetism,
the wave equation, spin-$s$ fields, fluids (simple, complex, or
even dissipative (\cite{LMR}, \cite{GL}), stressed solids, 
gravititation\footnote{The situation for the gravitational case
is more complicated than suggested here, because of the 
diffeomorphism freedom.} (\cite{CB}, \cite{FR}), 
etc. -- all have hyperbolizations\footnote{There is one example
of a system that seems ``physically reasonable", and yet does not
admit a hyperbolization --- the conducting fluid.  The fields
are $(F_{ab}, u^a, \rho, \mu)$, where $F_{ab}$ is the (skew)
Maxwell field, $u^a$ is the (unit) 4-velocity, $\rho$ is the
(positive) mass density, and $\mu$ is the charge density. 
The equations are Eqns. (2)-(5), with the Lorentz-force term
$F^a{}_bJ^b$ inserted on the right in (5) and the current-source
term $J_a$ inserted on the right in (2); together with charge-conservation,
$\nabla^aJ_a = 0$.  Here, the charge-current
is given by $J_a = \mu u_a + \sigma F_{ab}u^b$ (these two terms
representing, respectively, the charge-current due to bulk motion of the
charges in the fluid and the conductivity-current), where
$\sigma$ is some fixed constant (the electrical conductivity).
In fact, it does not seem to be possible to achieve an
hyperbolization for this system by any obvious modifications, e.g.,
including the twist of the velocity as an additional field.  I do
not know whether this system satisfies the conclusion of the 
initial-value formulation (later); nor, if it does, what the causal 
cones are.  It would be of interest to understand what is happening
with this system.}.  
Note that the existence of a hyperbolization depends only on the 
$k^{Aa}{}_{\alpha}$ of the system, and not on its $j^A$. 

Fix a first-order, quasilinear, system of partial differential
equations, (1), and a hyperbolization, $h_{A\beta}$, for
that system.  Fix also a point $p$ of $M$, and a value for the
fields, $\phi^{\alpha}$, at that point.  By the {\em causal cone}, 
$\cal{C}$, of
this system (at this point and for this field-value) we mean the set
of all tangent vectors $\xi^a$ at $p\in M$, such that
$\xi^an_a > 0$ for every $n_a$ for which $n_a h_{A\beta}k^{Aa}{}_{\alpha}$
positive-definite.  We note that, quite generally, $\cal{C}$ is
a nonempty, open, convex cone of tangent vectors at the point $p$
of $M$.  Note that $k^{Aa}{}_{\alpha}$ and $h_{A\beta}$ in general
depend on the choice of field-value at $p\in M$; and so, therefore, 
does the causal cone at $p$. 

For many examples --- Maxwell's equations, the Klein-Gordon equation, the
neutrino or Dirac equations, the spin-s field equation and Einstein's
equation --- the causal cone is precisely the future light cone.
[A choice of ``future" was singled out by the choice of sign in
the hyperbolization $h_{A\beta}$.] For the case of a simple fluid, the 
situation is a little more complicated.  First,
there is no hyperbolization at all unless $dp/d\rho > 0$.  When
this inequality is satisfied, there is a hyperbolization, and 
the corresponding causal cone $\cal{C}$ is given by a cone of
vectors $\xi^a$ satisfying $u_a\xi^a < 0$ and 
$(g_{ab} + (1-dp/d\rho)u_au_b)\xi^a\xi^b > 0$.  This will be 
recognized as the ``sound cone" of the fluid --- i.e., the
set of directions in space-time whose speed, measured with respect
to the fluid 4-velocity $u^a$, is less than the sound
speed $v$, given by $(v/c)^2 = dp/d\rho$.   Note that
when $v < c$, then all the vectors in $\cal{C}$ are timelike;
whereas when $v > c$ there are necessarily some spacelike vectors in
$\cal{C}$.  We emphasize that all of these remarks about fluids ---
the existence of a hyperbolization and of the causal cone --- 
apply equally well in the case of a 
subluminal and a superluminal sound speed.  The case of a stressed
solid is very similar to that of a fluid.  There are functions of
state that must be fixed to specify equations for this system, and
these functions give rise to acoustical-wave speeds associated with
the solid.   For appropriate functions of state, there exists a
hyperbolization.   These acoustical-wave speeds may be greater than
or less than the speed of light $c$; and the corresponding causal
cones may lie outside or inside the light-cones, respectively.  

Note that all of the above --- the existence of a hyperbolization,
and the corresponding causal cone --- depend only on the coefficient
$k^{Aa}{}_{\alpha}$ in the system (1) of equations, and not on
the $j^A$.  This observation has important implications.  For instance,
it follows immediately (from the corresponding facts for the wave
equation) that the mass-$m$ Klein-Gordon equation has a hyperbolization, with
causal cone the light cone.  Note that this holds even for the
Klein-Gordon equation with the ``wrong" sign for the term $m^2\psi$.
It follows further that the act of turning on an interaction between
two systems does not change the hyperbolization nor the causal
cones of the system.  
 
The causal cones deserve their name.  Roughly speaking, any 
first-order, quasilinear system is capable of sending signals 
only within its 
causal cones.  This remark is made precise by the initial-value
formulation for such systems, which we now describe.  

Fix a first-order, quasilinear system of partial differential
equations, together with a hyperbolization, $h_{A\beta}$, for that
system.  Let $S$ be a 3-dimensional submanifold of the manifold
$M$, and let there be given fields $\phi_o^{\alpha}$ on $S$.  We call
this $(S, \phi_o)$ {\em initial data} for our system provided
that, at each point of $S$, the closure of the causal cone at that
point lies entirely on one side of $S$.  [This means, in other
words, that a normal $n_a$ to $S$ is such that $n_ah_{A\beta}
k^{Aa}{}_{\alpha}$ is positive-definite.]  Note that these
causal cones in general {\em depend} on the field $\phi_o$ on $S$:
Thus, changing only $\phi_o$, keeping the submanifold $S$ fixed,
may render $(S, \phi_o)$ no longer initial data.  We now have:
\\

\noindent {\bf Initial-Value Formulation.}  Fix a first-order,
quasilinear system of
partial differential equations, a hyperbolization $h_{A\beta}$ for 
this system, and initial data $(S, \phi_o)$ for this system.  Then

   1.  There exists a neighborhood $U$ of $S$ in $M$, together with
fields $\phi^{\alpha}$ in $U$, such that i) $\phi|{}_S = \phi_o$, and
ii) $\phi$ satisfies\footnote{We are ignoring constraints in this
discussion.  The actual theorem guarantees only that the $\phi$ satisfy 
the ``$h_{A\beta}$"-part of Eqn. (1).  The rest of that equation is
also, ultimately, satisfied, but it is handled in a different manner.} 
Eqn. (1) in $U$. 

   2.  The field $\phi$ at a point $p$ of $U$ depends only on the
initial data in a region $A$ of $S$ such that $p$ is in the
domain of dependence of $A$.    
\\

\noindent  Item 1 guarantees a solution $\phi$ of our equation,
in some neighborhood of $S$, that reproduces the given initial
data.  [Note that we do not guarantee a solution in {\em all} of
$M$:  The fields may, for example, so evolve to become singular.]  
In item 2, $p$ lying in the {\em domain of dependence} of $A$
means that every curve $\gamma$ in $U$ that ends at $p$ and
that has its tangent vector at each of its points lying within
the causal cone at that point, meets $A$ once and only once.
By ``depends on" in
item 2, we mean that two sets of data on $S$, provided they agree
within $A\subset S$ yield the same field $\phi$ at $p$. 
Thus, item 2 asserts, in essence, that solutions of Eqn. (1) 
send signals that always lie within the causal cones:  If region
$A\subset S$ records every such signal that reaches point $p\in U$,
then that record, stored on $A$, determines completely what is
happening at $p$.  Note that the domain of dependence
in general depends on the field-values, since the causal cones do. 

The assertion above, in particular, guarantees an
initial-value formulation even for a fluid with superluminal sound speed.

Here is an example of these ideas.  Let us combine two systems,
turning on some interaction between them.
Let the respective coefficients in their differential equations 
be $k^{Aa}{}_{\alpha}$ and $k'^{A'a}{}_{\alpha'}$, and let their 
respective hyperbolizations be $h_{A\beta}$ and $h'_{A'\beta'}$.  
Then $(h_{A\beta}, h'_{A'\beta'})$ is our candidate for 
an hyperbolization of the combined system.  It
automatically satisfies the symmetry condition; and it satisfies
the positive-definiteness condition provided that, at each point,
there is some covector $n_a$ such that both $n_ah_{A\beta}k^{Aa}{}_{\alpha}$
and $n_ah'_{A'\beta'}k'^{A'a}{}_{\alpha'}$ are positive-definite.
Suppose that this condition holds, so we indeed have a hyperbolization 
for the combined system.  
Then the set of covectors $n_a$ that give positive-definiteness
for this hyperbolization is precisely the set of $n_a$ for which
$n_ah_{A\beta}k^{Aa}{}_{\alpha}$
and $n_ah'_{A'\beta'}k'^{A'a}{}_{\alpha'}$ are both positive-definite.
It follows that the causal cone of the combined system is the
convex hull of the causal cones of the two individual systems ---
i.e., the set of all sums of the form $\xi^a + \xi'^a$, with
$\xi^a$ in $\cal{C}$ and $\xi'^a$ in $\cal{C}'$.  
This is what, in light of he statement above, we would expect physically.
Signal propagation,
for the combined system, is in those space-time directions 
obtained by taking sums of the signal-propagation
directions for the two systems separately.

To summarize, each first-order, quasilinear system of partial
differential equations -- provided that system has a hyperbolization
--- carries within itself its own initial-value formulation.  And,
as a part of that
formulation, the system carries its own causal cones for signal 
propagation.  These cones are inherent in the structure of the 
equation itself, i.e., they do not necessarily require that 
there be fixed any outside fields.  We may combine such systems
--- and turn on interactions between systems --- and when we do
so the causal cones also combine, in the way we expect physically. 

This formulation manifests what might be called a democracy of causal 
cones.  All systems, and their cones, are on an equal footing:  No 
one set of fields, or one set of causal cones, has priority over any 
others.  

\section{Special Relativity}

We now turn to the special theory of relativity.  This theory
involves, of course, a flat metric, $g_{ab}$, on the space-time
manifold $M$.  

We first note that ``flat metric" can be restated in terms of 
a first-order, quasilinear system of partial differential equations.  The
fields consist of the metric $g_{ab}$ and the derivative operator $\nabla_a$;
and the equations are 
\begin{equation}
\nabla_a g_{bc} = 0, \ \ 
R_{abcd} = 0,
\end{equation}
where $R_{abcd}$ is the Riemann tensor.   Note that this system
is indeed first-order and quasilinear.  It is true that 
the solutions of this system are rather uninteresting 
dynamically, e.g., 
because they are all ``locally identical".  Indeed, this is probably
the reason why one does not normally think of special
relativity in terms of two fields satisfying a system of
partial differential equations. 

We claim, next, that the system (6) admits a hyperbolization\footnote{
Here, again, we are ignoring the diffeomorphism freedom, which, again, 
does not materially impact the present considerations.}!  Indeed,
this is a consequence of the existence of a hyperbolization
for Einstein's equation, in light of the fact that the Einstein
system is merely a subset 
of the system (6).  The causal cones for the system of
special relativity are, of course, the null cones of the metric
$g_{ab}$.

So, we may adopt the perspective that ``special relativity" is merely
one more first-order, quasilinear system of partial differential
equations admitting a hyperbolization.  It is just one more physical
theory, not dissimilar from the theory of electromagnetism or the
theory of a simple fluid.  Like all such systems, special relativity carries 
with it, by virtue of the structure of its equations, causal cones.  
Some systems,
such as that of electromagnetism, share those cones with special
relativity; while other systems, such as that of a fluid, do not.
But each system --- special relativity included --- looks to its 
own causal cones --- to its own system of partial differential equations
--- for the propagation of signals within that theory.  

In short, the causal cones of special relativity, from this perspective,
have no special place over and above the cones of any other
system.  This is democracy of causal cones with a vengeance.  This, 
of course, is not the traditional view.  That view --- that
the special-relativity causal cones have a preferred role in
physics --- arises, I suspect, from the fact  that a number of other 
systems --- electromagnetism, the spin-$s$ fields, etc --- employ precisely
those same cones as their own.  And, indeed, it
may be the case that the physical world is organized around
such a commonality of cones. 
On the other hand, it is entirely possible that there exist any
number of other systems --- not yet observed (or maybe they have
been!) ---  that employ quite different sets of
causal cones.  
And the cones of these ``other systems" could very well lie outside
the null cones of special relativity, i.e., these systems could
very well manifest superluminal signals.  None of this would
contradict our fundamental ideas about how physics is structured:
An initial-value formulation, causal cones governing signals, etc.  

To illustrate these points,
let us return to the example of the system consisting
of special relativity together with a fluid manifesting superluminal
sound signals.  This is a completely viable system of partial
differential equations.  It has, in particular, an initial-value
formulation.  Initial data must be specified on a 3-dimensional
surface $S$ that is spacelike (as determined by the metric
$g_{ab}$ of special relativity), {\em and} is such that each fluid
sound-cone lies on one side of $S$.  These data then evolve,
producing fields $g_{ab}, \nabla_a, u^a$, and $\rho$
within the domain of dependence of those initial data, as determined 
by the causal
cones of the combined system.  This system does not differ, in any
essential way, from the system of special-relativity-electromagnetism.
In particular, this system is ``Lorentz-invariant", at least in
the sense that any $g_{ab}$-preserving diffeomorphism on $M$ sends
solutions of the fluid equations to new solutions.  
There is nothing peculiar happening here.

We discussed at the beginning two ``concrete" arguments in support of
the idea that it is appropriate to take the nonexistence of
superluminal signals as a fundamental principle of physics.  It is 
instructive to return to those arguments, in light of the 
discussion above.

The first involved the difficulty in generating superluminal signals.
One example of this was the problem of accelerating a particle to a
superluminal speed, in light of the mass-increase formula,
$m = m_o/\sqrt{1-v^2/c^2}$.  From the present perspective, this
``difficulty in generating superluminal signals" merely reflects
the fact that we are trying to create such signals using fields
(such as electromagnetism, etc) that take as their
causal cones the null cones of special relativity.  The 
``mass-increase formula" is now seen, not as a general
property of {\em all} particles, but rather as a property only
of particles constructed from such fields.  Were
there other fields --- with other causal cones --- and were we
able to construct particles from these fields, then those
particles would manifest ``mass increase" 
by quite a different formula.  These newly constructed particles 
could then be used to
transmit superluminal signals.  Of course, such particles
could not be used to achieve a signal-speed greater than that dictated 
by the causal cones of those new fields.

The second argument involved the grandfather paradox.  Let us first
consider the arrangement in which the two pipes have already been
set up, with one moving rapidly past the other.  This is, presumably,
a solution of the special-relativity-superluminal-fluid system.
But it has closed causal curves (via, of course, the causal cones
of this combined system).  It follows that this arrangement cannot
be in the domain of dependence of any surface, i.e., it cannot
be ``predicted", via the initial-value formulation, from any initial
data.  Our observer will object at this point, claiming that he
can ``build" precisely this arrangement:  
First lay out the two pipes of fluid parallel to each other and at rest,
and then accelerate one of those pipes along its length.  Our response to
this objection is the following.  We grant that the observer can 
set up those initial conditions (the two pipes
at rest).  But the issue of what happens from those initial
conditions must be determined by evolving, from those initial data,
the differential equations for the system.  [Presumably,
we would include also within this system the fields describing the
observer, and the initial conditions would reflect that observer's 
resolve to get
the one pipe moving.]  Whatever results from these data and these 
equations is what results.  But we know that ---
whatever it turns out to be --- the result of this evolution 
will {\em not} consist of 
the two pipes moving in the prescribed manner.  Probably,
it will be difficult to include in the system interactions that
will allow the observer to move the fluid around in the manner
he wishes --- for example, the fluid may interact back with the
observer, preventing him from manipulating that fluid in the
desired manner; or, because of its equations, the fluid might
respond to such manipulation in an unexpected manner.  It is also
possible that the fluid solutions themselves might become 
singular when the fluid is pushed too hard. 

This circumstance is not as strange as it might seem at first glance.
Indeed, it arises all the time in physics.  Suppose, for example,
that an observer reported that he planned to build, and then
use as a signalling device, certain electromagnetic fields specified
on a timelike surface.   We would certainly insist on knowing the details of
how this is to be done.  We would grant this observer the power to
set up some initial conditions for the electromagnetic field (on a
{\em spacelike} surface).  But we would then insist that the final
field configuration is, not what the observer wills it to be, but
rather what follows, evolving these data via Maxwell's equations.
If the observer can achieve the desired field-configuration in this way,
we will accept it; if not, we will not.  And, in a similar vein, there exist
solutions of Einstein's equation in general relativity that 
manifest closed causal curves.  But we do {\em not}, in
light of this circumstance, allow observers to build time-machines 
at their pleasure.  Instead,
we permit observers to construct initial conditions --- and then
we require that they live with the consequences of those conditions.
It turns out that a ``time-machine" is never a consequence,
in this sense,
of the equations of general relativity, in close analogy with the
situation in the special-relativity-superluminal-fluid
example above.

To summarize, from the present viewpoint the problems associated with 
superluminal signals do not seem nearly as severe as they did at first
glance.   

Here, in any case, is another perspective on special relativity.
The theory emerges as just one more physical system.  It consists, 
just like the others, of certain fields subject to a certain system of
first-order, quasilinear partial differential equations.  The
causal cones of special relativity (which reflect the speed of light) 
have no special significance over the causal cones of any of the
many other such systems in physics.  I am not sure that this is 
the right perspective --- or even whether ``right" makes much sense in 
this context.  But I would suggest that this viewpiont has something 
to offer.

\end{document}